\documentclass[conference,compsoc]{IEEEtran}
\usepackage{amsmath}
\usepackage{graphicx}
\ifCLASSOPTIONcompsoc
  \usepackage[nocompress]{cite}
\else
  \usepackage{cite}
\fi
\ifCLASSINFOpdf
\else
\fi
\ifCLASSOPTIONcompsoc
 \usepackage[caption=false,font=footnotesize,labelfont=sf,textfont=sf]{subfig}
\else
 \usepackage[caption=false,font=footnotesize]{subfig}
\fi
\begin{document}
%
\title{Robustness enhancement of cloud computing network based on coupled networks model}

\author{\IEEEauthorblockN{Zibin Su}
\IEEEauthorblockA{School of Computer Science\\
Beijing University of Posts and\\Telecommunications\\
Beijing, China\\
Email: su\_zibin@163.com}
\and
\IEEEauthorblockN{Jing Yuan}
\IEEEauthorblockA{School of Computer Science\\
Beijing University of Posts and\\Telecommunications\\
Beijing, China\\
Email: yuanjing\_sky@163.com}}


%


\maketitle

\begin{abstract}
  As a novel technology, cloud computing attracts more and more people including
  technology enthusiasts and malicious users. Different from the classical
  network architecture, cloud environment has many its own features which make
  the traditional defense mechanism invalid. To make the network more robust
  against a malicious attack, we introduce a new method to mitigate this risk
  efficiently and systematically. In this paper, we first propose a coupled
  networks model which adequately considers the interactions between physical
  layer and virtual layer in a practical cloud computing environment. Based on
  this new model and our systematical method, we show that with the addition of
  protection of some specific nodes in the network structure, the robustness of
  cloud computing's network can be significantly improved whereas their
  functionality remains unchanged. Our results demonstrate that our new method
  can effectively settle the hard problems which cloud computing now is facing
  without much cost.
\end{abstract}


%
\IEEEpeerreviewmaketitle

\section{Introduction}
The flourish of cloud computing technology nowadays is largely due to its
outstanding features, on-demand self-service, resource pooling, rapid elasticity,
etc.[1] All users in cloud environment share a public pool of configurable and
virtualized computing resources, such as CPUs, disks or network. Users can
easily scale-up or scale-down their cloud resources according to their real
time demands. For example, before the landing of NASA's Curiosity rover, IT
engineers are allowed to deploy as many servers running on the AWS (Amazon Web
Services) cloud as they need. Then, when they are done, they may shut down
additional servers to avoid paying for those resources [2]. Besides, the
centralization of servers makes cloud computing technology more environment
friendly and energy saving. Compared to setting up their own data center,
individuals and enterprises are now becoming more favorable to deploy their
businesses on cloud [3] [4]. Cloud computing has leveraged users from hardware
requirements, while reducing overall client side requirements and complexity [5].

As the fast growth of cloud computing, security issues are considered as the
obstacles on the highway, which largely hinder the big enterprises' wills of
porting their business from traditional data center to cloud. Apparently, the
security of cloud computing seems to be improved due to the centralization of data
and increased security-focused resources [6]. The fact, however, is that the
security of cloud computing now is considered still in infancy [7], especially
the network security which faces many new challenges.

Generally, to protect an enterprise network against cyber-attack, we
traditionally adopt network security devices such as firewalls, DMZ hosts or
intrusion detection systems (IDS) [8]. These traditional network defense
strategies, however, can not be applied to cloud computing environment
adaptively due to not only the attacks can rise internally but also the
dynamic and elastic features of cloud computing [9]. To settle such problems,
new methods are proposed continually in the past years, such as distributed
cloud intrusion detection model proposed by Irfan Gul and M. Hussain[10],
integrating an IDS into cloud computing environment proposed by Claudio
Mazzariello, Roberto Bifulco and Roberto Canonico[11] or control the
inter-communication among virtual machines method proposed by Hanqian Wu,
Yi Ding[12], etc.

These novel methods, however, merely try to reinforce cloud computing's
internal network via porting traditional network defense means. Such methods
are not only unsystematic but also impossible to implement when the scale of
physical hosts reaches at least half million [13]. Besides, once the
cyber-attack causes some VMs overloaded which in turn causes physical hosts
which they reside overloaded, all the services on this overloaded physical
hosts will be affected or even be corrupted. Moreover, due to the logical
coupling between VMs in a common virtual sub-network, for example, the coupling
relationship between load-balancers and servers or between servers and databases,
disasters will spread dramatically and then quickly collapse a large part of
cloud network.

In this paper, we first propose a new two layers model to describe the cloud's
complex internal network with the full consideration of the interactions between
physical and virtual networks. Based on this model and complex network theory,
a novel solution is introduced to systematically and globally settle such a
problem that the traditional network defense strategies are no longer suitable
for cloud. This solution can make the whole network in cloud computing
environment more robust to resist the malicious cyber attacks and to maintain
the infrastructures as operatively as possible, even before collapsing.

\section{Avalanche Effect in Cloud Computing's Cyber Attack}

\begin{figure}[!t]
  \centering
  \includegraphics[width=8cm]{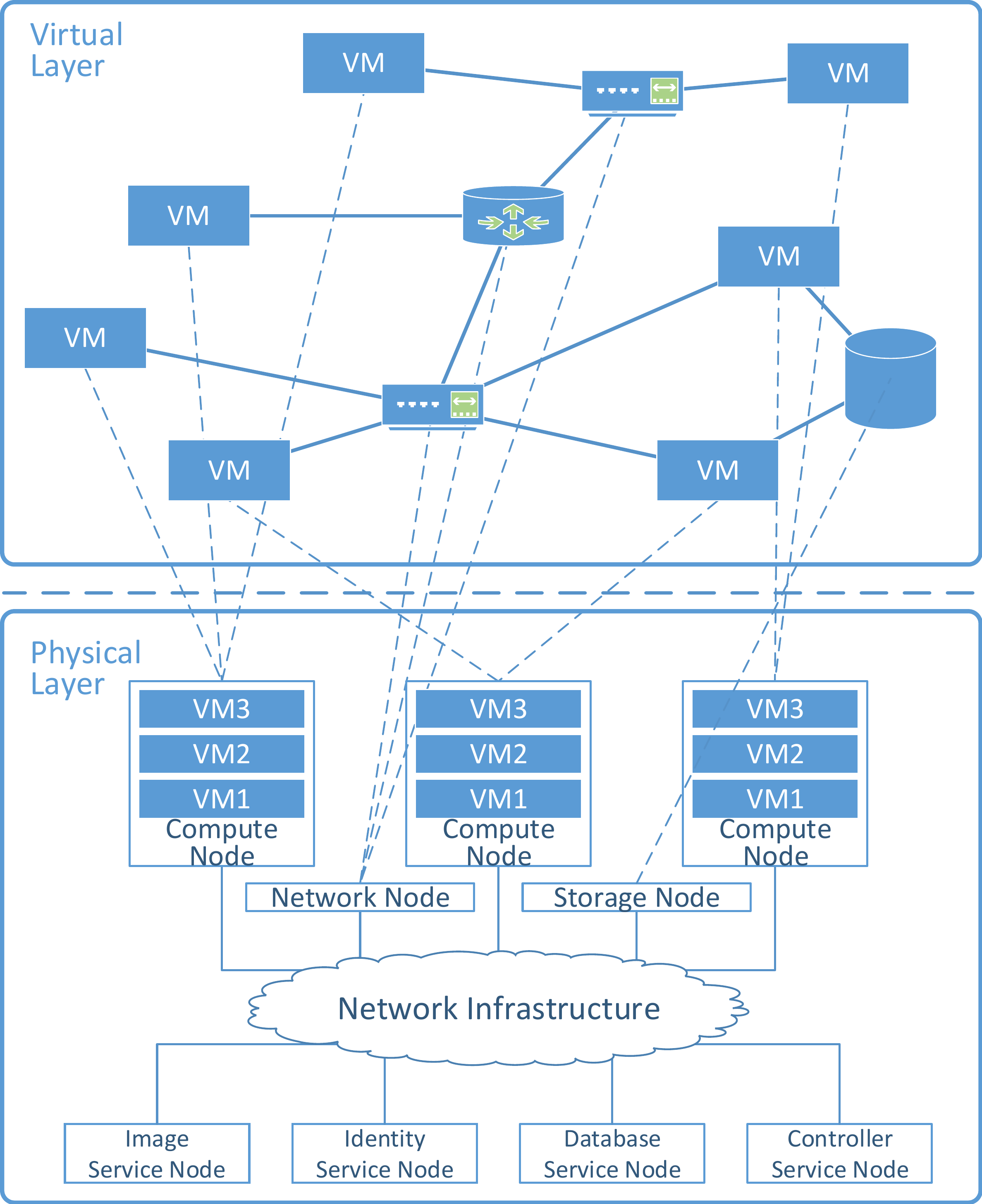}
  \caption{The relationship between virtual and physical layer in the network of cloud computing. On the physical layer, servers or other physical devices are applied to various virtualization technologies to service for the virtual layer. The virtual layer, on the other hand, doesn't know the exist of physical layer. Various virtual sub-networks coexist with each other in the virtual layer, each of them is deployed by cloud user for different purpose.}
\end{figure}

Different from the traditional networks, network in cloud computing environment
can be divided into two layers: the virtual layer and the physical layer. The
physical layer contains chunks of physical network facilities, such as switchers,
routers, severs or other common network devices. The virtual layer, however, is
built on the physical layer and is implemented via various virtualization
technologies, such as container technologies, virtual machine technologies or
software define network technologies [14]. All these virtual resources, such as
VM instances, distributed databases or distributed storage, run on physical hosts
and are inter-connected via virtual networks which also run on some physical
hosts [15]. Fig 1 shows the relationship between virtual and physical layers.

Due to the sharing of physical resource pool, crash of one VM instance can cause
other VM instances on the same physical host to collapse. Furthermore, because
of the logical coupling of different components in a sub-virtual network, such
collapsing may spread along different paths on both physical layer and virtual
layer. This process is the avalanche effect in cloud computing's cyber attack.
Fig. 2 shows the avalanche process when only one VM instance in the network is
attacked by malicious hacker. According to the complex network theory, such
avalanche effect can ruin a network rapidly, even the scale of a network is
really large [16].

\begin{figure}[!t]
  \centering
  \includegraphics[width=8cm]{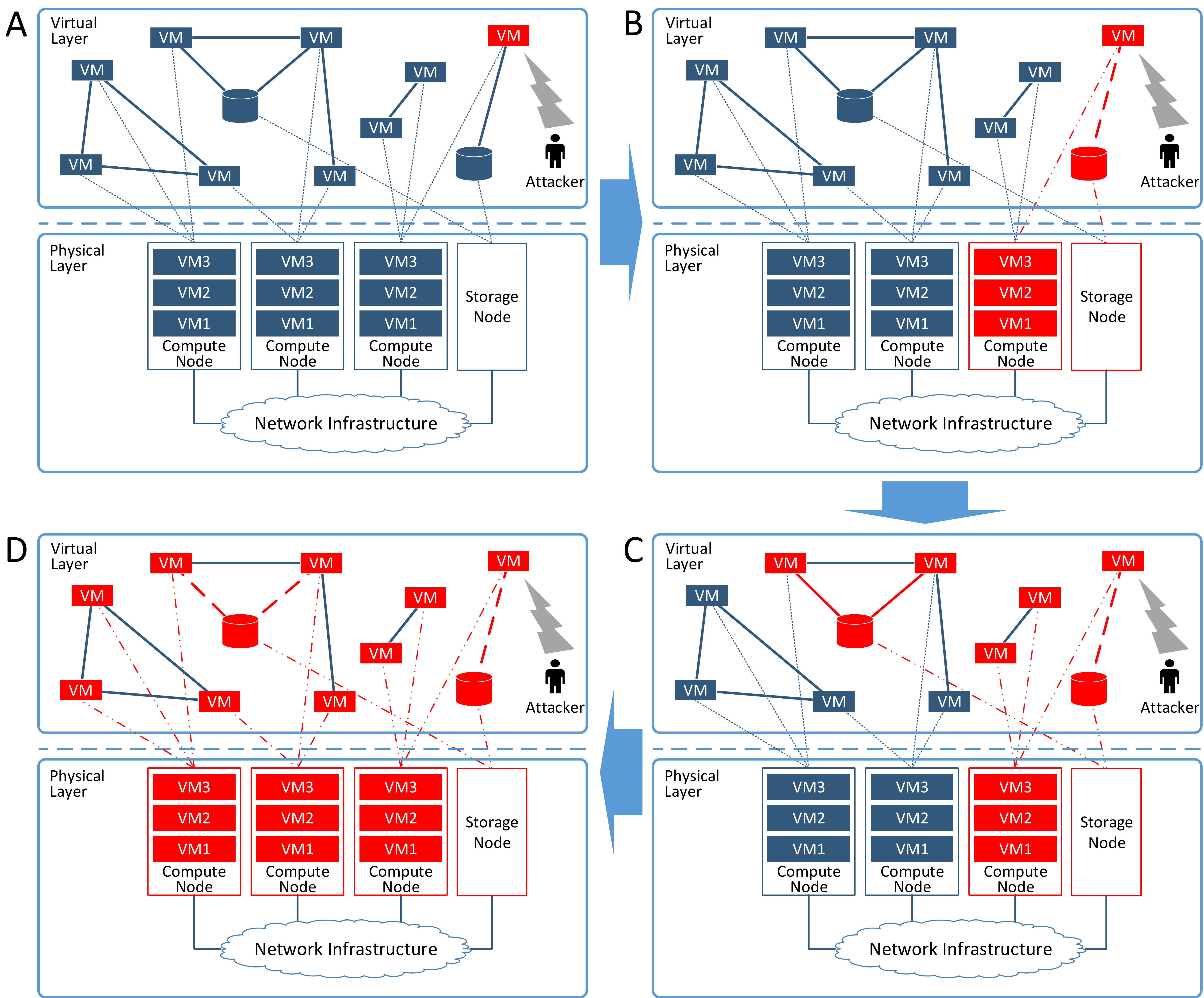}
  \caption{The avalanche process when cloud computing's is under a cyber attack. Initially, in (A), only one VM is attacked. Due to coupling of physical layer and virtual layer, however, other VMs which beside the ruined ones can be affected or even be collapsed with a probability P. This process will continue until no more new ones. After this avalanche process, in the worst case, all VMs in the network is in an invalid state (D).}
\end{figure}

\section{Model of Cloud Computing's Network}

\subsection{Modeling Virtual and physical layer}
In the real cloud environment, VM instances and other virtual components compose
the virtual sub-network which represents a full functional application, such as
a web application or scientific computing platform. The whole virtual layer
consists of various virtual sub-networks that have different scales. In this
article, we use scale-free network to model such a virtual sub-network, due to
the fact that many kinds of practical computer networks, including the internet
and local area networks, are all scale-free [17]. Here, we suppose that the
distribution of virtual sub-virtual networks with different scale obey the
Power Law distribution, i.e. the larger the virtual sub-network is, the litter
it appears [18]. After this, we can deploy these virtual components onto physical
machines randomly.

The modeling process can be divided into two steps: 1) Generate various
scale-free networks whose scale distribution obey the Power Law distribution.
Each vertex in this network represents a virtual component, such as VM instance
on which runs different services. 2) Create the two-layers model by adding
physical vertices into the network and randomly add edges between these physical
vertices and the vertices in virtual layer.

\subsection{Simplify the model}
To apply complex network methods to the model, we need to simplify the two
layers' model into single layer. Ignoring the specific functions of facilities
in the two layers, all virtual or physical components in cloud can be treated as
a vertex in network, and inter-connections between vertices can be abstracted
as edges. Due to that virtual network is built on the physical network, edges
between vertices on physical layer can be omitted. So, this two-layer network
can be further simplified to a single layer network as we can see from Fig. 3
[19]. This simplification reduces the complexity of analysis and makes it
possible to use the mature complex network theory and tools.

\begin{figure}[!t]
  \centering
  \includegraphics[width=8cm]{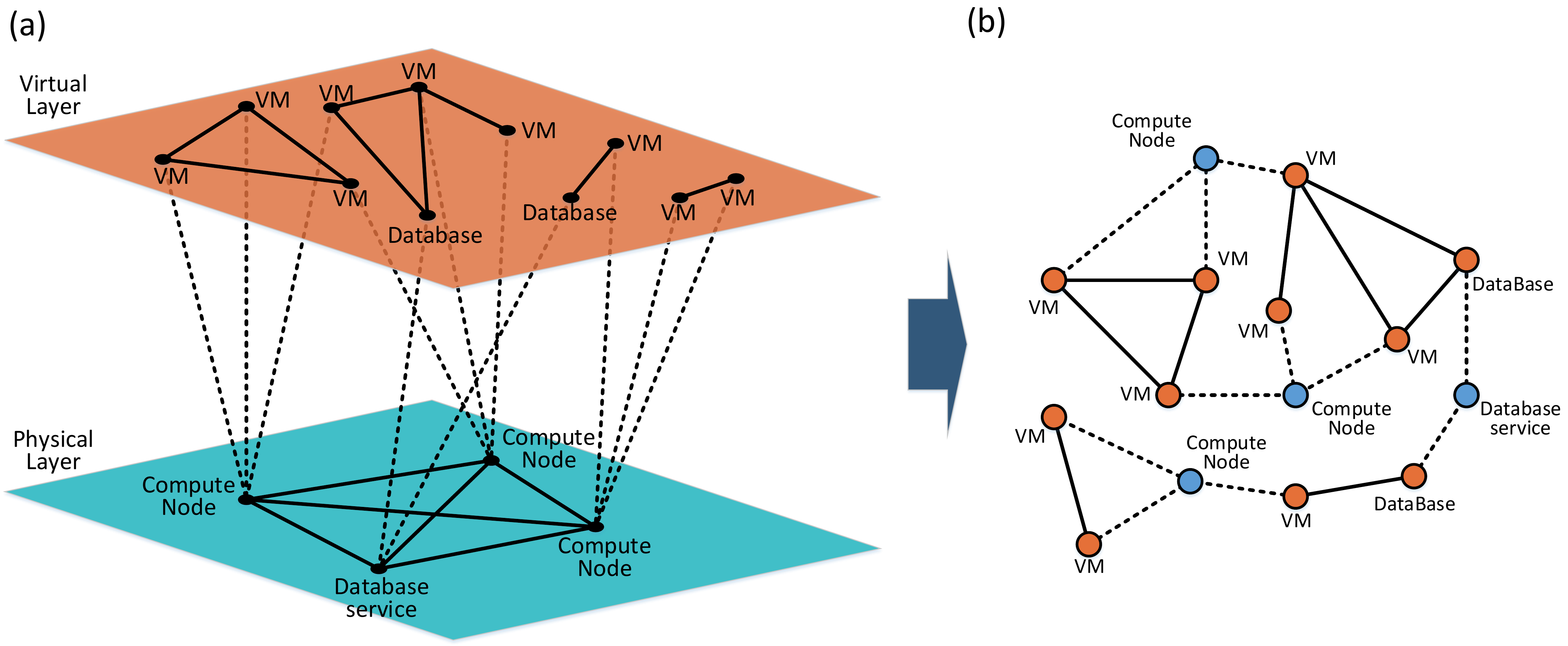}
  \caption{Simplify two-layers network to single-layer network. This simplification ignores the connection between the physical vertices due to that the virtual layer network is built on the physical layer network. After this simplification, for the one-layer network, we can employ the classical complex network analysis method to settle the problem that we have encountered.}
\end{figure}

Then, we come to analyze the robustness of this system. We usually use the size
of giant component after initially removing a fraction $q$ of nodes to measure
the robustness of a network. First, we consider the situation in which no immune
nodes are set up to guarantee the function of the whole network. Bond percolation
process can be a great tool to model the dynamic process in the system. Edges are
occupied only when the end nodes of the edges are not initially removed and both
the end nodes are not infected (node $i$ are infected with probability $1-P_{imu}(i)$,
we will discuss it later). We define $\pi_{i}(s)$ as the probability that node $i$
belongs to a small clusters of exactly $s$ nodes. Since the network is sparse
enough, we can assume that the network topology is locally tree-like. This means
that in the limit of large network size an arbitrarily large neighborhood around
any nodes takes the form of a tree, then the calculation using message-passing
algorithms can give a good approximation of the clusters.

Assuming that the networks to be locally tree-like, according to Brian Karrer and
M. E. J. Newman's recent theory [22], $\pi_{i}(s)$ can be write as:

\begin{equation}
  \pi_{i}(s)=\sum_{\{s_{j}:j \in N_{i}\}}\bigg[\prod_{j \in N_{i}} \pi_{i \gets j(s_{i})} \bigg]
  \delta \Big(s-1, \sum_{j \in N_{j}} s_{j}\Big)
\end{equation}

Where $\delta(a,b)$ is the kronecker delta which is defined as follows:

\begin{equation}
  \delta (a,b) = \left\{ \begin{array}{ll}
    0 & \textrm{$a-b = 0$}\\
    1 & \textrm{$a-b \neq 0$}
    \end{array} \right.
\end{equation}

We can now introduce a probability generating function $G_{i}(z)=\sum_{s=1}^{\infty}\pi_{i}(s)z^s$
, whose value is given by [22]:

\begin{equation}
  \begin{split}
  G_{i}(z)&=\sum_{s=1}^{\infty}z^s\sum_{\{s_{j}:j \in N_{i}\}}\Big[\prod_{j \in N_{i}}\pi_{i \gets j}(s_{j}) \Big]
  \delta(s-1,\sum_{j \in N_{i}}s_{j})\\
  &=z\prod_{j \in N_{i}}\sum_{s_{j}=0}^{\infty}\pi_{i \gets j}(s_{j})z^{s_j}
  \end{split}
\end{equation}

We can simplify the equation as [22]:

\begin{equation}
  G_i(z)=z\prod_{j \in N_i} H_{i \gets j}(z)
\end{equation}

Where $\prod_{j \in N_i} H_{i \gets j}(z)=\sum_{s=0}^{\infty}\pi_{i \gets j}(s)z^s$.

To calculate $H_{i \gets j}(z)$, we note that $\pi_{i \gets j}(s)$ is zero if
the edge between $i$ and $j$ is unoccupied (with probability $1-p_{i \gets j}$)
and nonzero otherwise ($p_{i \gets j}$), which means that $\pi_{i \gets j}(0)=1-p_{i \gets j}$
in which:

\begin{equation}
  p_{i\gets j}=(1-\eta)^2P_{imu}(j)
\end{equation}

Where $\eta$ stands for the fraction of nodes initially removed. And for $s\ge1$:

\begin{equation}
  \begin{split}
  \pi_{i \gets j}(s)=p_{i\gets j}\sum_{\{s_k:k\in N_{j\backslash i}\}}\Big[\prod_{k\in N_{j\backslash i}}\pi_{j \gets k}(s_k) \Big]
  \delta(s-1, \sum_{k \in N_{j\backslash i}}s_k)
  \end{split}
\end{equation}

Where the $N_{j\backslash i}$ denotes that the set of neighbors of $j$ without $i$.
Substituting this equation into the definition of $H_{i\gets j}(z)$ above,
we then find that:

\begin{equation}
  H_{i\gets j}(z)=1-p_{i\gets j} + p_{i\gets j}z\prod_{k\in N_{j\backslash i}}H_{j\gets k}(z)
\end{equation}

Then the expected fraction $S$ of the network occupied by the entire percolating
cluster is given by the average over all nodes:

\begin{equation}
  S=\frac{1}{n}\sum_{i=1}^{\infty}\Big[1-G_i(1) \Big]=1-\frac{1}{n}\sum_{i=1}^n\prod_{j\in N_i}H_{i\gets j}(1)
\end{equation}

Setting $z=1$ in equation (7) we have:

\begin{equation}
  H_{i\gets j}(1)=1-p_{i\gets j} + p_{i\gets j}\prod_{j\in N_{j\backslash i}}H_{i\gets j}(1)
\end{equation}

We can calculate the size of the remaining greatest connected component of the
networks, i.e. the percolating cluster by solving this equation.

\subsection{Vertices with immune ability}
Some nodes in this network may have the immune ability against malicious
attacks due to that they are well protected by   some virtual network security
equipments which are deployed by professional network administrators. Usually,
large corporations have enough money and awareness to employ professional
security counselors and managers to protect their IT facilities (physical or
virtual) from cyber-attacks. According to this common sense, in our model,
vertices in a large virtual sub-network will have a high probability to avoid
crash when they are attacked by hackers. The immunity probability $P_{imu}$
of a specific node V can be calculated as:

\begin{equation}
  P_{imu} = 1-\frac{S}{T} \cdot C
\end{equation}

where $S$ is the number of vertices in a virtual sub-network which node V
belongs to and $T$ is the number of vertices in the whole virtual layer.
Coefficient $C$ stands for that even a virtual sub-network is well protected,
it is also possible to be ruined inevitably by some cases.

\subsection{Solution}
To enhance the robustness of cloud computing network, we may place some key
vertices behind virtual network security components, such virtual firewall,
virtual IDS etc. [21] In our virtual layer model, we don't take account of
virtual network security components due to that they are transparent to the
application users and can't be attacked directly. Virtual network security
components can be deployed rapidly and conveniently without much more
consumptions.  The key vertices which are selected to protect are that have the
highest degrees in the network. Usually, vertex has high degree somehow means
that they are important or even crucial.

\begin{figure}[!t]
  \centering
  \includegraphics[width=8cm]{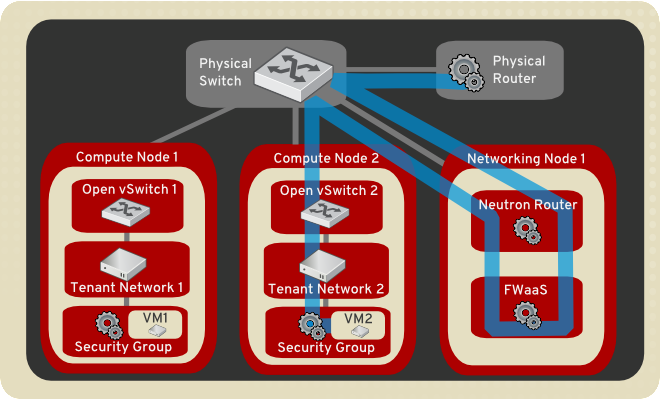}
  \caption{This figure is referenced from Firewall-as-a-Service (FWaaS) overview of Openstack Cloud Administrator Guide [20]. In this figure, VM2 is protected by a virtual firewall. In our model, we can place such key nodes behind the security components.}
\end{figure}

To simulate the crashing process, initially, we randomly remove some vertices
from the network to simulate that some VMs are ruined. Then all the vertices
which are the neighbors of crashed nodes are affected. Due to that each node in
the network has its own immune coefficient which we have mentioned before, the
neighboring nodes may survive and avoid crashing during the process. These new
crashed nodes in turn affected their own neighbors. This process will continue
until the system reaches a stable state that no more vertices are affected.  In
each spreading step, we use the number of nodes in the largest connected cluster
to represents the current state of network.

Based on the model we have discussed before, we now consider the situation with
immune nodes which are totally immune to the infections and will never collapse
with some protection. In this paper, we select the nodes with greatest degrees
as the immune nodes. As the introduce of immune nodes into the system, there will
be some changes for $p_{i \gets j}$.

\begin{equation}
  p'_{i \gets j} = (1-\eta)^2\bigg(1-\Big[1-P_{imu(j)} \Big](1-P_{j\in B})\bigg)
\end{equation}

Where $P_{j\in B}$  stands for the probability that $j$ is in the selected group
of immune nodes. If the immune nodes are randomly selected, $P_{j\in B}=\eta$
for all $j$ and $\eta$ is the fraction of protected nodes. It is obvious that
$p_{i\gets j} < p'_{i\gets j}$, thus the expectation of size of giant component
will be greater. In this paper, we selected the nodes with greatest degrees as
immune nodes. Thus $P_{j\in B}$  will be a function of its degree and $\eta$ .
And if $j$ is in the group with greatest degrees, the effect of protecting this
node will be greater .Therefore the network will be more robust. 

\section{Results}
\begin{figure*}[!t]
  \centering
  \includegraphics[width=17.6cm]{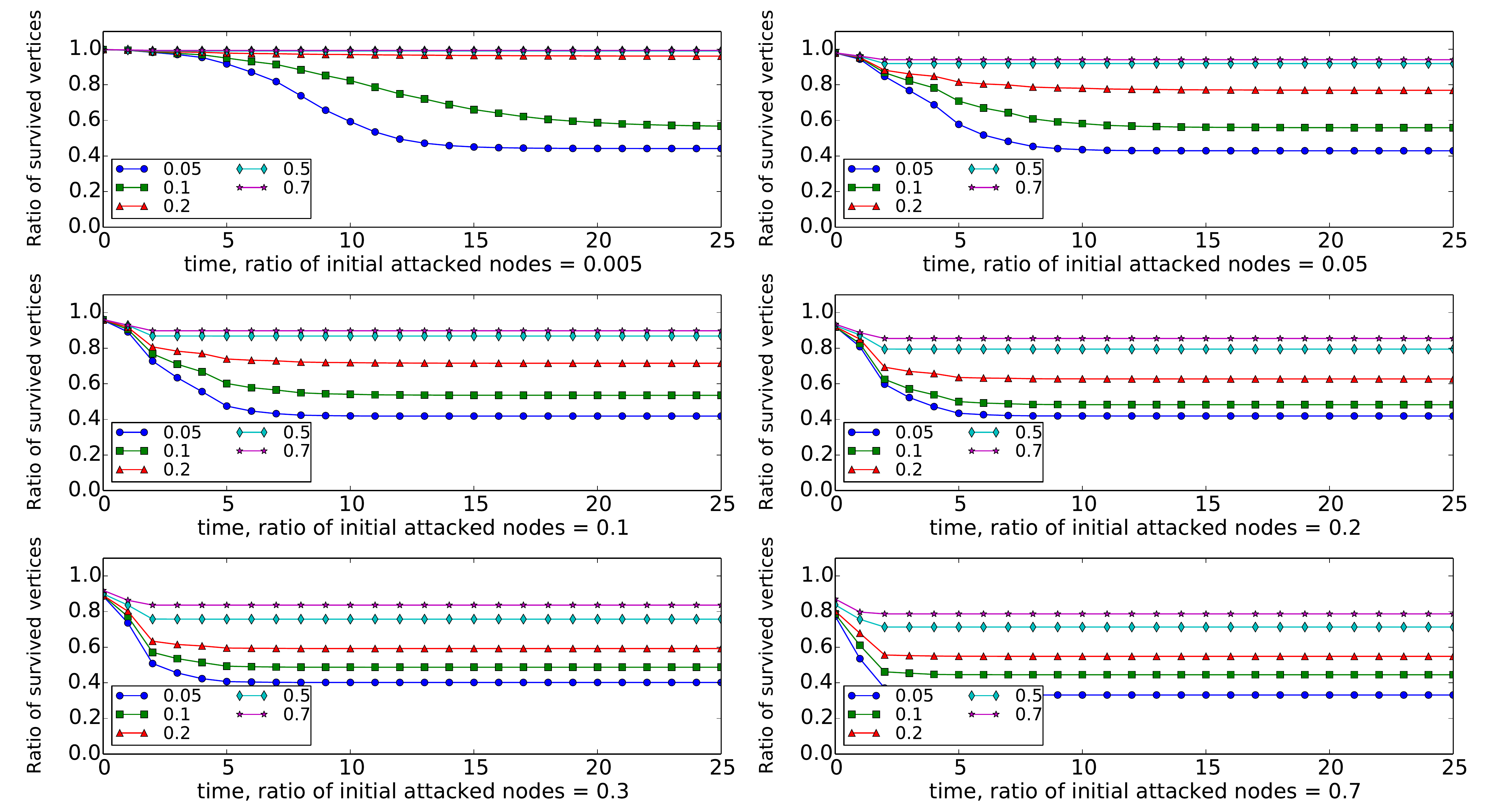}
  \caption{With different number of immune nodes, the ratio of survived nodes in largest connected cluster are significantly different. That only protect 0.05\% nodes in the network will keep the ratio of survived nodes over 50\%. If the ratio of immune nodes increased to 20\%, in some common cases (5\% nodes are attacked initially), ratio of survived nodes will even over 95\%}
\end{figure*}

From what we have discussed in the above sections, we can conclude that as long
as the immune nodes are added into the network, the probability of existing
larger cluster is improved as well. To verify the robustness improvement after
applying our novel method to cloud computing's network, we have simulated this
avalanche process with different ratio of initial immune nodes and initial
attacked nodes. We use 5000 physical hosts with 10 VM instances running on each
of them to simulate the attack process. Here, we assume that the largest scale
of virtual sub-network contains at most 500 VM instances.

The results in Fig. 5 show that the number of key nodes that have the ability
to resist the cyber-attack will finally affect the robustness of the whole
network, and the initial number of attacked nodes also affects the network's
robustness. As the ratio of initial attacked nodes increase, the number of
survived nodes in the largest connected cluster decrease accordingly. Also,
with different ratio of protected nodes, the robustness (measured by the number
of nodes in the largest connected cluster) of network varies significantly.
The more the nodes are protected, the higher the robustness is. Fig. 5
demonstrates that if we only select 5\% (2500 VMs, 250 physical hosts) key nodes
to give the ability to resist the cyber-attack, the ratio of final survived nodes
to the total nodes can increased over 40\% or even 70\%. Also, if we protect 20\%
key nodes, this ratio will stably over 60\% and in some optimistic cases it will
over 90\% (0.5\% nodes initially be attacked).

In practice, benefited from the elastic and dynamic features of cloud computing,
nodes can by rapidly protected by virtual network security devices on demand.
Besides, the SDN technology has the ability to detect the real time topology and
to re-calculate the degree of all nodes in network rapidly. So that, when we
detect the change of network, no matter physical or virtual, we can re-select
the key nodes (nodes have the highest degree) and protect those new key nodes
by the virtual network components to obtain the immunity in a short period.

\section{Conclusion}
In summary, we have introduced a novel method based on complex network theory
that can significantly improve the robustness of cloud computing's network to
defense malicious attacks with low costs. Our approach shows that with a
reasonable protection of some key nodes in the network, significant gains can be
achieved for the robustness while the network's functional topology keep
unchanged. This result reveals the fact that instead of deploying security
equipment on each rack, protecting the key nodes with virtual network security
components is more efficient, economic and energy-saving. The applications of
our results are imminent on one hand to guide the improvement of the existing
cloud computing networks but also serve on the other hand to design future cloud
infrastructures with improved robustness.

\end{document}